\documentclass[a4paper,fleqn,usenatbib]{mnras}

\usepackage[T1]{fontenc}
\usepackage{ae,aecompl}

\usepackage{graphicx}	
\usepackage{amsmath}	
\usepackage{amssymb}	
\usepackage{xspace}

\title[New outburst from the luminous SSS$_1$ in NGC~300]{New outburst from the luminous supersoft source SSS$_1$ in NGC~300 with periodic modulation}

\author[S. Carpano et al.]{
S. Carpano,$^{1}$\thanks{E-mail: scarpano@mpe.mpg.de}
F. Haberl,$^{1}$
C. Maitra$^{1}$
\\
$^{1}$Max-Planck-Institut f\"{u}r extraterrestrische Physik, Giessenbachstra{\ss}e 1, 85748 Garching, Germany}

\date{Accepted 2019 October 15. Received 2019 October 14; in original form 2019 September 5}

\pubyear{2019}

\newcommand{\xmm}{{\it XMM-Newton}\xspace}

\newcommand{\ergs}{erg s$^{-1}$}

\begin{document}
\label{firstpage}
\pagerange{\pageref{firstpage}--\pageref{lastpage}}
\maketitle

\begin{abstract}The nearby galaxy NGC~300 is hosting two luminous transient supersoft X-ray sources with bolometric luminosities above $3\times 10^{38}$ \ergs, assuming simple black-body spectra with temperatures around 60--70\,eV. For one of these, SSS$_1$, a periodic modulation of 5.4\,h was observed in an \xmm observation from 1st of January 2001 lasting 47\,ks, but not visible 6 days earlier when the luminosity was higher. We report here the detection of a new outburst from this source, which occurred during two more recent \xmm observations performed on 17 to 20 December 2016 lasting for 310\,ks. The luminosity was similar as in December 2000, and the 0.2$-$2.0 keV light curve revealed again a periodic modulation, with a period of 4.68$\pm$0.26\,h, significant only in the first of the two observations. Taking into account the large uncertainties (the 2001 period was re-estimated at 5.7$\pm$1.1\,h), the two values could be marginally compatible, and maybe associated with an orbital period, although the signal strength is highly variable. Thanks to the new long exposures, an additional absorption feature is now visible in the spectra, that we modelled with an absorption edge. This component decreases the bolometric luminosity below $3\times 10^{38}$\,\ergs\ and would therefore allow the presence of a white dwarf with a mass close to the Chandrasekhar limit. The system was found in outburst in 
1992, 2000, 2008, and 2016 suggesting a possible recurrence period of about 8 years. We discuss viable models involving white dwarfs, neutron stars or black holes. \end{abstract}




\section{Introduction}
Ultraluminous supersoft sources (ULSs) are X-ray sources that have bolometric luminosities exceeding the Eddington limit for a 1.4\,M$_\odot$ compact object when modelled by black-body spectra with temperatures typically below 100\,eV.
The nature of these sources is believed to be different from the classical SSSs whose bolometric luminosities, in the range of $10^{36} - 2 \times 10^{38}$ \ergs, are explained by steady nuclear burning of hydrogen accreted onto white dwarfs \citep{vandenHeuvel1992}.

Many of these ultraluminous SSSs have been detected in nearby galaxies: two in M 101 \citep{DiStefano2003,Kong2004}, one in M 51 \citep{DiStefano2003}, one in M 81 \citep{Swartz2002, Liu2008}, one in the Antennae \citep{Fabbiano2003}, one in  NGC~4631 \citep{Carpano2007},  two in NGC~300 \citep{Kong2003,Carpano2006} and one in NGC~247 \citep{Jin2011}. Eleven more candidates have been spotted by \cite{Sazonov2017}, based on their hardness-ratios (0.25--2\,keV to 0.25--8\,keV flux ratio >0.95).

Despite having similar observed characteristics (super-Eddington luminosities and thermal spectra with temperatures below 100\,eV),  ULSs can harbour a heterogeneous class of objects with different behaviours. The companion star could be of low or high mass, and the compact object can be either a neutron star, stellar mass black hole or even an intermediate mass black hole. In the case of M81 ULS1, the unexpected discovery of relativistic jets suggests the presence of supercritical accretion around a black hole with optically thick outflows \citep{Liu2015}.

For luminosities around $3\times 10^{38}$ \ergs, the presence of a very massive white dwarf radiating at its Eddington limit could still be invoked. If the source presents as well recurrent outbursts, one could be dealing with a recurrent nova. Indeed, those systems host white dwarfs with masses > 1.2\,M$_\odot$ \citep[with a mean value expected at 1.3\,M$_\odot$,][]{Shara2018} which might increase as matter continues to pile up with a mass accretion rate estimated in the range of 10$^{-7}-10^{-8}$\,M$_{\odot}$ yr$^{-1}$ \citep{Shara2018}. The accumulated material will eventually start a thermonuclear explosion that makes the nova eruption, with recurrence time scales of less than a century.  The 10 known Galactic recurrent novae can be subdivided in three groups according to the orbital period: very short periods (several hours) for systems with low-mass main sequence companion star, periods around one day with subgiant companions, and very long periods above one year with red giant companions \citep{Schaefer2010}. Recurrent novae, like other novae, undergo supersoft X-ray  phases which are delayed by several days/weeks with respect to the optical outburst. This phase can only be observed once the ejected matter becomes transparent to soft X-rays, at a time that is generally defined as the \emph{turn-on time} (t$_\textrm{on}$). \cite{Henzel2014} showed that for the SSSs in M31, this timescale can greatly vary from days to years and is likely correlated with the \emph{turn-off time} (t$_\textrm{off}$), time at which hydrogen burning stops (although sources with $\textrm{t}_\textrm{off}< \textrm{t}_\textrm{on}$ can't be observed). For systems found with t$_\textrm{on}$> 1 year, it was suggested by \cite{Pietsch2005} that these could host a recurrent nova for which the optical outburst responsible for the soft X-ray emission was not detected.

NGC~300 is a face-on nearby spiral galaxy, located at 1.88\,Mpc \citep{Gieren2005} with low Galactic foreground column density of $N_\textrm{H} = 3.6\times 10^{20} \textrm{cm}^{-2}$ \citep{Dickey1990}. Two very luminous supersoft sources have been discovered so far in the galaxy, both transients. The first one, reported by \cite{Kong2003}, and called SSS$_1$ in \cite{Carpano2006}, is a recurrent source, with a bolometric luminosity close to $10^{39}$ \ergs, assuming a pure black-body spectrum, and a temperature of  kT$\sim$ 60\,eV. The source appeared for the first time as a ULS in a ROSAT observation of 1992, and then in two \xmm observations on the 26 December 2000 and 1st of January 2001. In the second observation, lasting $\sim$13\,h, the source was found in a low state with a flux dropping by a factor of 9 and the light curve revealed a periodicity of 5.4\,h. The source was again found in outburst in a Swift observation of 2008 \citep{Kong2008}. So far, a periodic flux modulation was only found for one other ULS in NGC~4631, with a 4\,h period from an \xmm observation lasting $\sim$15\,h \citep{Carpano2007}. The second luminous and transient source in NGC~300, called SSS$_2$, was found in two subsequent \xmm observations from 22 May and 25 November 2005 where the bolometric luminosity decreased from 8.1 to $2.2\times 10^{38}$ \ergs, and during which the first source was off \citep{Carpano2006}.

We report in this article the detection of a new outburst of SSS$_1$, during two consecutive and very long \xmm observations of NGC~300 spanning over 310\,ks in December 2016. In Sec.~\ref{sec:optical} we derive precise source coordinates using a Chandra observation and look for possible optical counterparts, while in Sec.~\ref{sec:timing} and Sec.~\ref{sec:spectrum} we analyse its X-ray light curve and spectrum, respectively. 
We then compare in Sec.~\ref{sec:compar} these recent observations with previous ones from \xmm, Chandra, Swift and ROSAT.
We discuss the nature of the source and give our conclusions in the last two sections.

\section{The optical counterpart}
\label{sec:optical}
The detection of an outburst in two Swift observations from 20 May and 4 June 2008, triggered a new Chandra observation (obsID: 9883) pointed on SSS$_1$ on the 8th of July 2008 for an exposure of 10\,ks with ACIS-S instrument in \texttt{VFAINT} mode. This allows the measurement of a more accurate source position. The \texttt{\textbf{ciao}} tool \texttt{celldetect}, provides refined coordinates of $\alpha_\text{J2000}=00^\text{h}55^\text{m} 10\fs{}98$ and $\delta_\text{J2000}=-37^\circ 38' 54\farcs 47$ with a statistical error of $0\farcs 05$, which is negligible with respect to the absolute astrometric accuracy of $\sim 0\farcs 6-0\farcs 8$. 

\begin{figure}
   \centering
   \resizebox{\hsize}{!}{\includegraphics{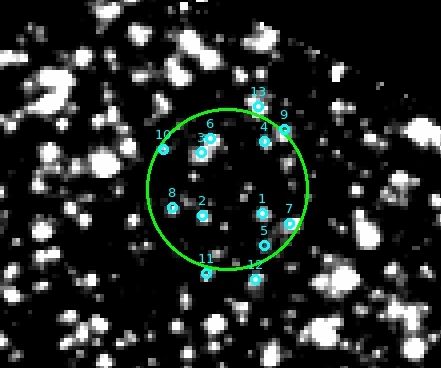}}
   \caption{Combined (F606W+F814W) HST ACS/WFC optical image around the Chandra position of SSS$_1$ (green circle with 1$''$ radius). The sources indicated by cyan circles in the region have magnitudes between 25 and 27 mag.}
              \label{fig:HST}             
\end{figure}

Figure~\ref{fig:HST} shows the HST ACS/WFC optical image combined for the F606W (broad V) and F814W (broad I) filters, recorded on the 2nd of July 2014 (downloaded from the MAST Portal\footnote{\url{https://mast.stsci.edu/portal/Mashup/Clients/Mast/Portal.html}}).  From the MAST HST source catalogue, we extracted the magnitudes for the sources located in the vicinity of SSS$_1$, indicated as small cyan circles. 13 sources, located within 1.2$''$ from the Chandra position, are included in the catalogue and have apparent V and I magnitudes between 25 and 27 mag, which, at the distance of NGC~300, would be translated to absolute magnitudes between $-$1.37 and +0.62 mag. These are marginally compatible with a massive companion star (early A stars or giants), while the fainter sources could only be low-mass stars. This means that, in the case of a neutron star or black hole compact object, the system can only be a Low-Mass X-ray binary, and that mass transfer is performed via Roche Lobe overflow.

\section{The December 2016 \xmm observation}

The supersoft source SSS$_1$ was again in outburst during two consecutive \xmm observations, 0791010101 and 0791010301, performed from the 17th to the 20th of December 2016 for 139 and 82\,ks respectively. 
We utilize data from the European Photon Imaging Cameras (EPIC) based on pn and MOS type CCD detectors \citep{2001A&A...365L..18S,2001A&A...365L..27T}.
The data were reduced following standard procedures using the \xmm SAS data analysis software version 17.0\footnote{Science Analysis Software (SAS): http://xmm.esac.esa.int/sas/} with the calibration files (CCFs) available in February 2019. For the source extraction and background regions we used a circle with a radius of 20$''$ around the Chandra coordinates, mentioned in Section\,\ref{sec:optical}, and of  35$''$ around $\alpha_\text{J2000}=00^\text{h}55^\text{m} 05\fs{}53$ and $\delta_\text{J2000}=-37^\circ 38' 52\farcs 04$, respectively.

\subsection{Light curve analysis}

\begin{figure}
   \centering
   \resizebox{\hsize}{!}{\includegraphics{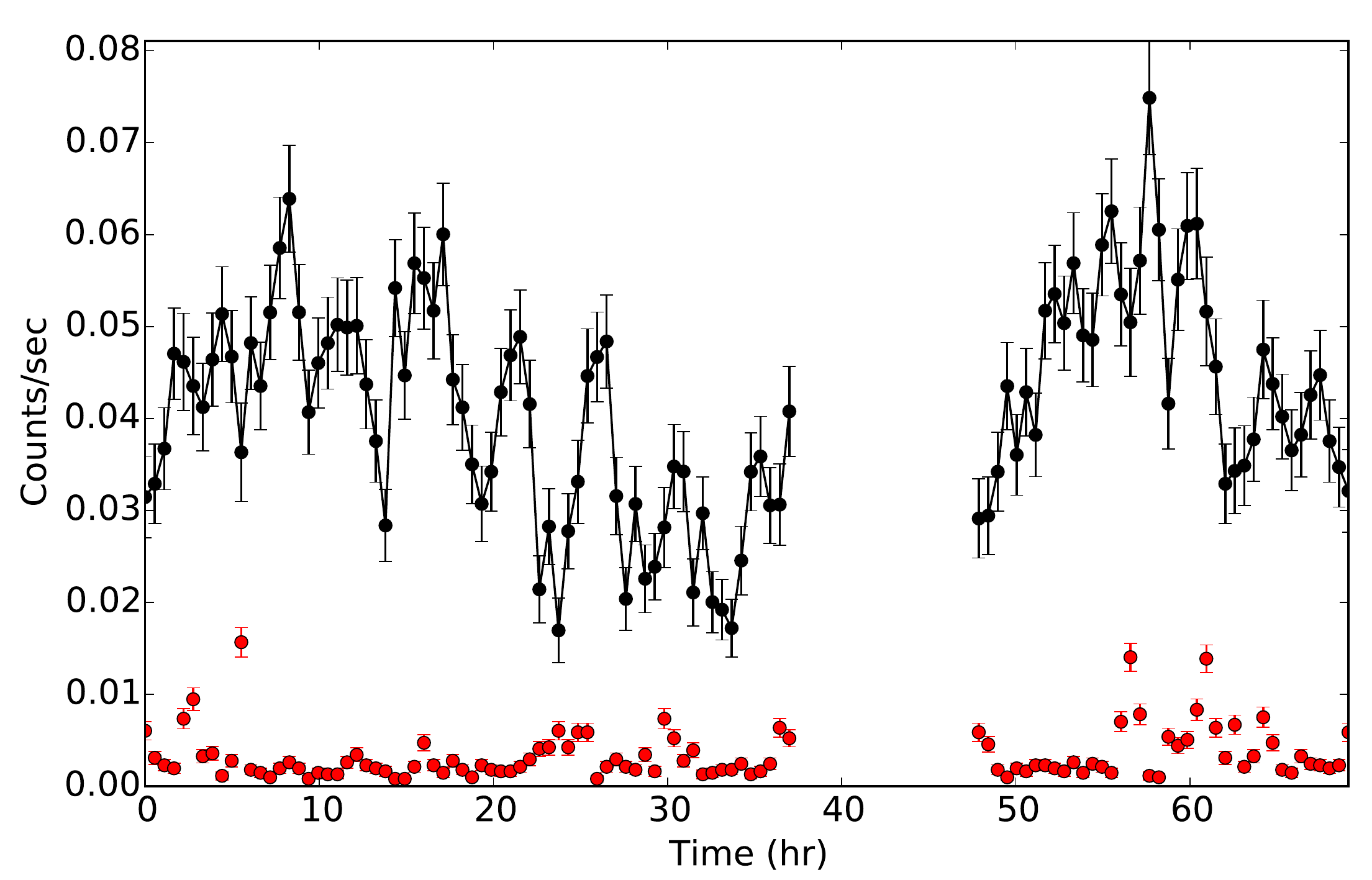}}
   \caption{\xmm EPIC (pn + MOS1 + MOS2) light curve of SSS$_1$ from Dec. 2016, rebinned at 2000\,s, extracted in the 0.2-2.0\,keV band. The red curve shows the corresponding background rate scaled for the respective areas.}
              \label{fig:LC}             
\end{figure}

\begin{figure}
   \centering
   \resizebox{\hsize}{!}{\includegraphics{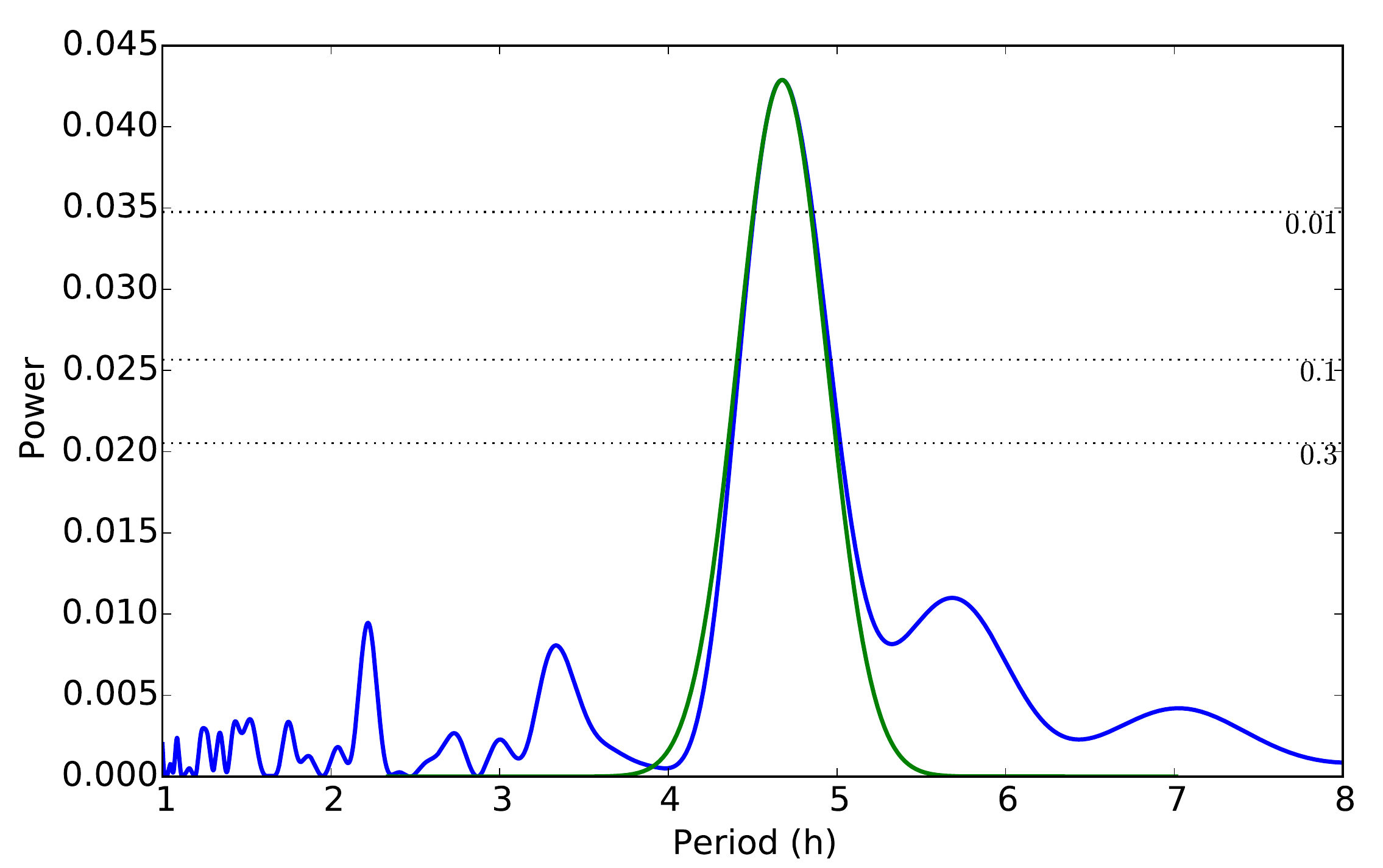}}
   \caption{Lomb-Scargle periodogram on the first part of the  light curve (obsid: 0791010101), binned at 100\,s. The best period is 4.68\,h and the 1-$\sigma$ uncertainty derived from the gaussian fit is 0.26\,h.}
              \label{fig:LS}             
\end{figure}

The background-subtracted EPIC (pn + MOS1 + MOS2) light curve binned at 2000\,s and extracted in the 0.2-2.0\,keV band is shown in Fig.~\ref{fig:LC} in black while the area scaled background light curve only is shown in red. We can see that during the first observation, 0791010101, a periodic modulation is clearly visible, while it is less obvious in the second one (0791010301).
To determine the period we performed a Lomb-Scargle periodogram analysis \citep{Lomb1976, Scargle1982} in the 1 to 8\,h period range, on the  light curve binned at 100\,s. The result for the first observation only is shown in Fig.~\ref{fig:LS}, with the best period being of 4.68\,h.  The uncertainty on the period is derived by fitting a Gaussian function to the highest peak and corresponds to its standard deviation (0.26\,h). The confidence levels (given at 68$\%$, 90$\%$ and 99$\%$) are derived using the block-bootstrap error as explained in \cite{Carpano2017}: we simulated 1000 light curves by randomly shuffling blocks from the light curve with lengths of $\sim$1000\,s. In the second observation no significant periodic signal was found, although one peak exists at 4.9$\pm$0.3\,h.

\begin{figure}
   \centering
   \resizebox{\hsize}{!}{\includegraphics{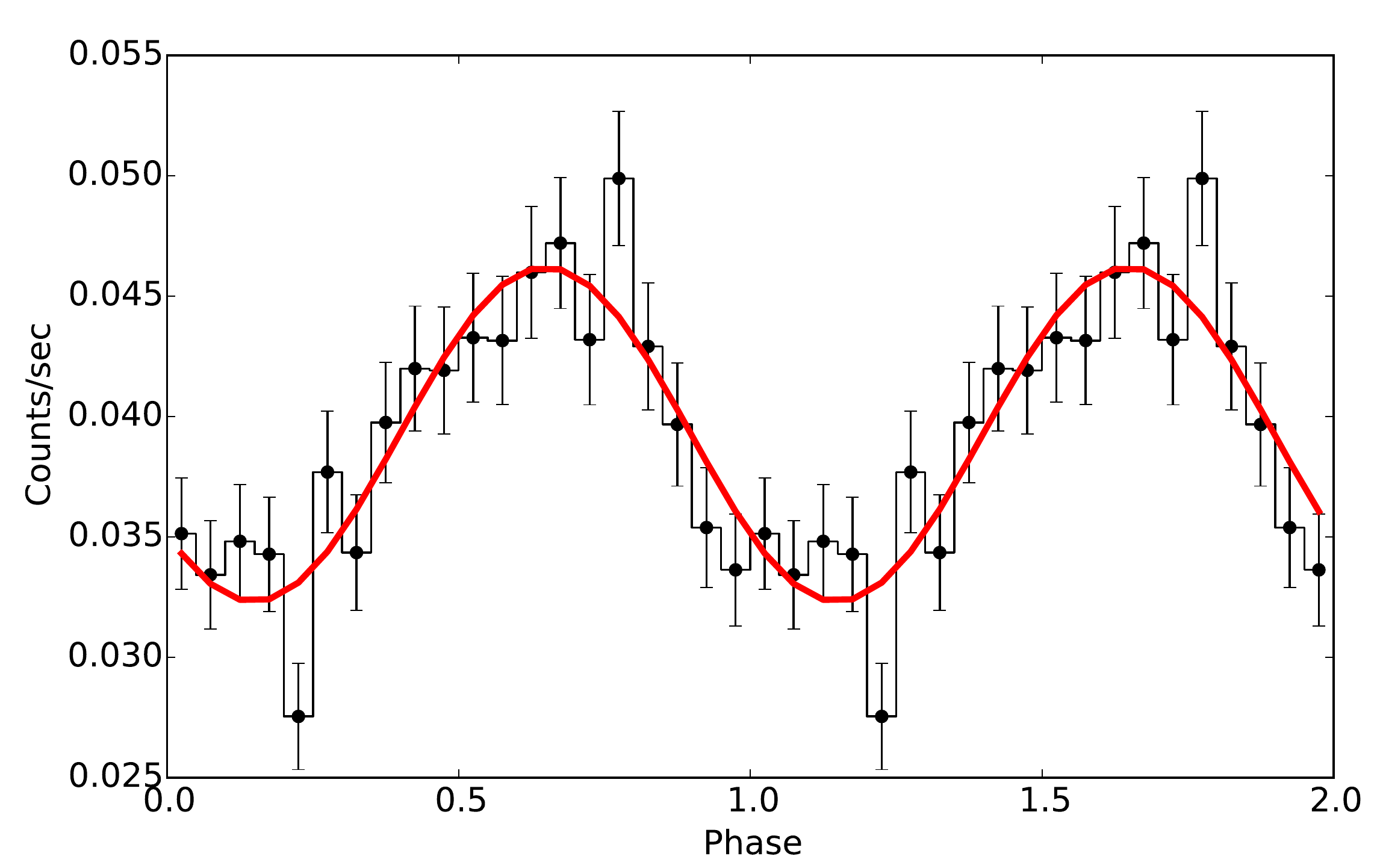}}
   \caption{Light curve of SSS$_1$ from the first observation folded at the best period of 4.68\,h, with best-fit sine function overlaid and with a mean count rate of 0.039\,cts\,s$^{-1}$.}
              \label{fig:fold}             
\end{figure}

\begin{figure}
   \centering
   \resizebox{\hsize}{!}{\includegraphics{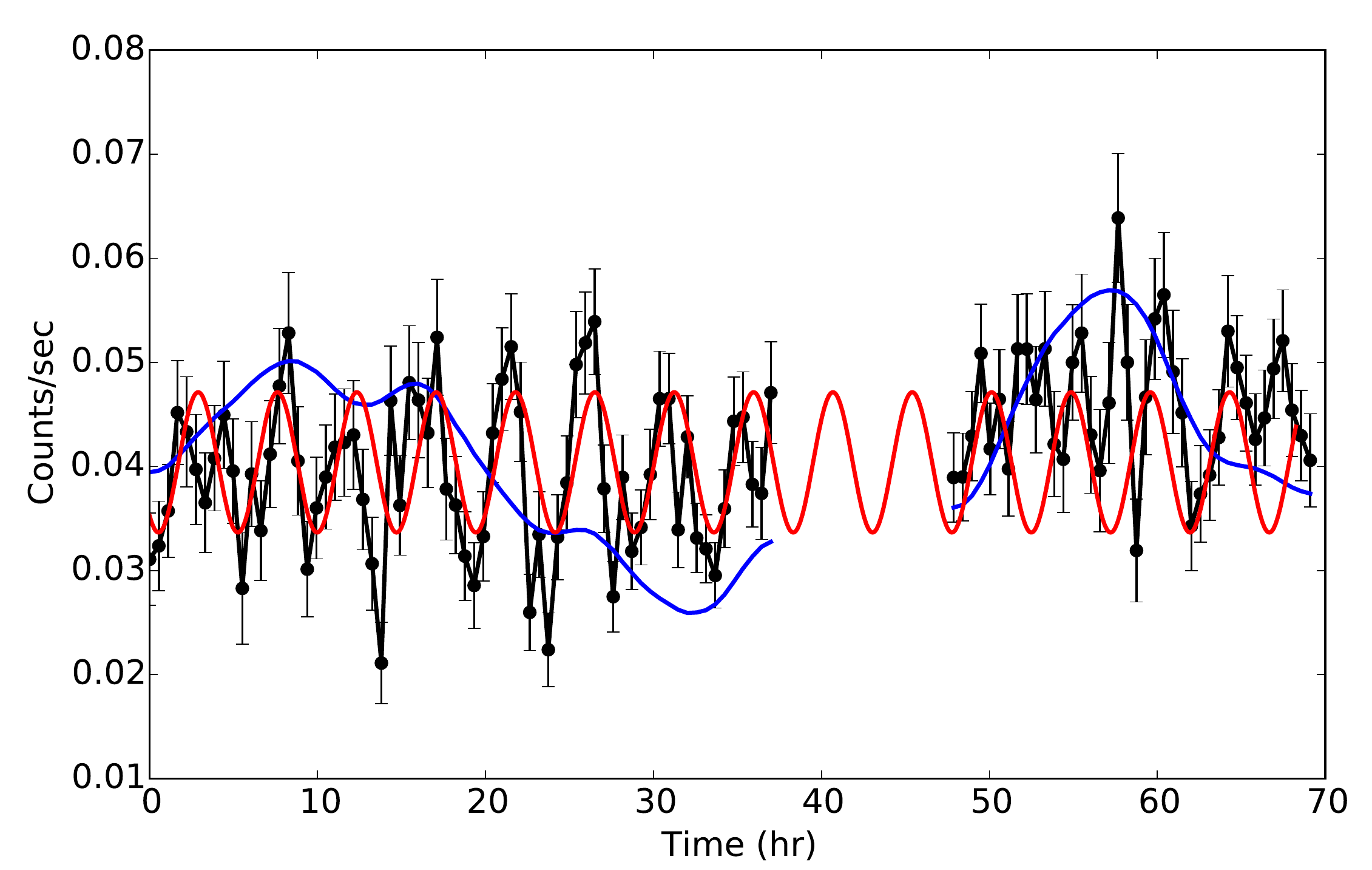}}
   \caption{Detrended light curve of SSS$_1$ (black) from Fig.~\ref{fig:LC}, with extrapolated sine function (red) and long-term trend (blue) overlaid.}
              \label{fig:sine}             
\end{figure}

The folded light curve of the first observation is shown in Fig.~\ref{fig:fold} together with the best fitted sine function, providing an amplitude of 0.0069\,cts\,s$^{-1}$ and a mean count rate of 0.039\,cts\,s$^{-1}$, leading to a modulation of 17.7$\%$. Fig.~\ref{fig:sine} shows the detrended light curve from Fig.~\ref{fig:LC} in black, with the long-term variation (smoothed light curve using a Hamming window function) in blue. The sine function derived from the folded light curve (Fig.~\ref{fig:fold}) is extrapolated over the entire light curve and is shown in red.

The period derived in this observation is marginally compatible with the January 2001 period, that we re-estimated to 5.7$\pm$1.1\,h. Since the signal strength largely varies from observation to observation it's unclear if the origin is related to the orbital motion. Another possible origin could be a super-orbital motion from a precessing warped disc  (although much longer periods are expected) while low-frequencies quasi-periodic oscillations (QPOs) should be of much shorter periods (tens of milliseconds to few hundred seconds).

\label{sec:timing}

\subsection{Spectral modeling}
\label{sec:spectrum}

\begin{figure}
   \centering
   \resizebox{\hsize}{!}{\includegraphics{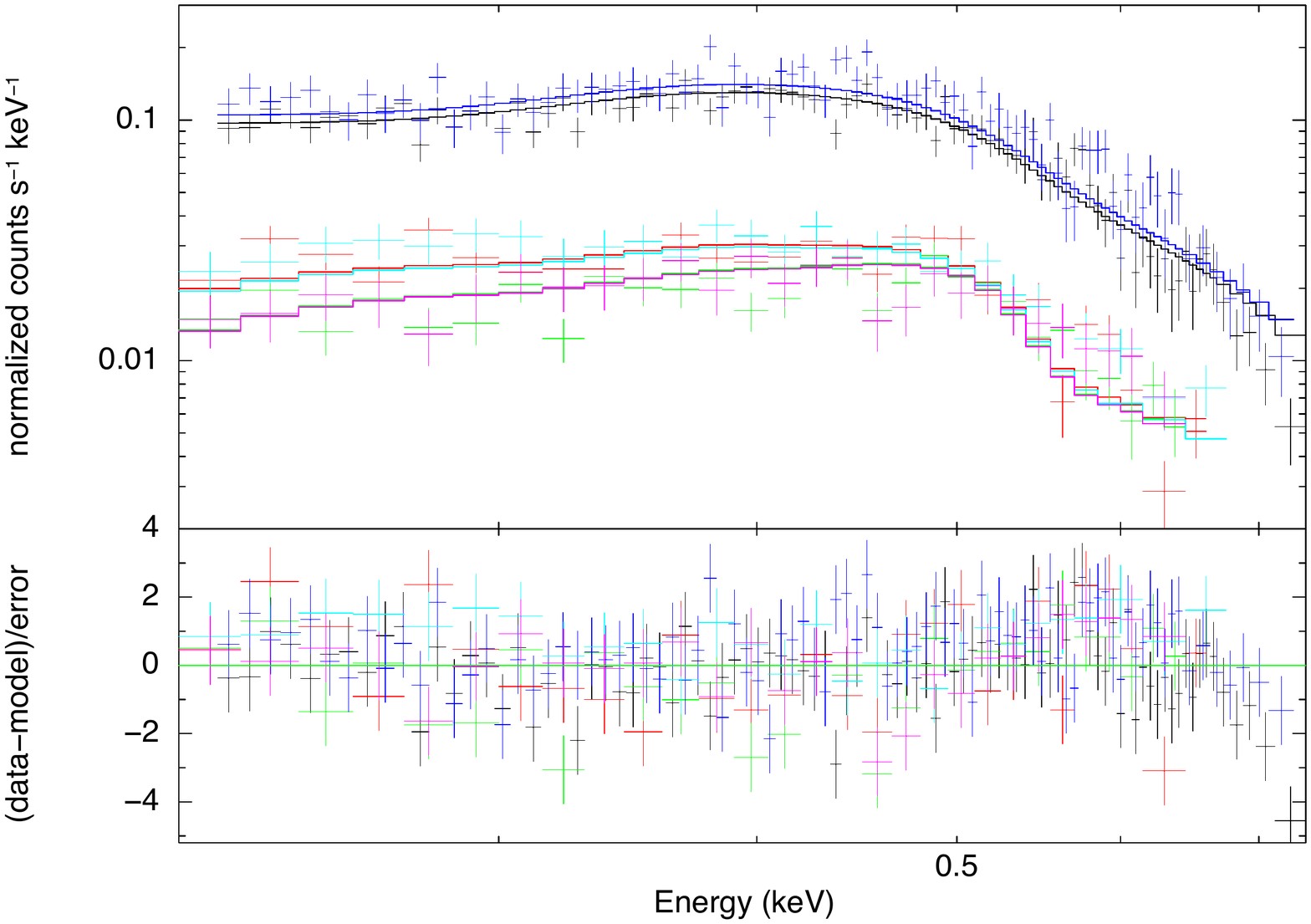}}
   \caption{Simultaneous spectral fit of SSS$_1$ from both observations 0791010101 (black: pn, red: MOS1, green: MOS2) and 0791010301 (blue: pn, cyan: MOS1, pink: MOS2), using an absorbed black-body model (\texttt{tbabs*bbody}), and the corresponding residuals.}
              \label{fig:spec}             
\end{figure}

We use \texttt{PyXspec} version 2.0.1\footnote{\url{https://heasarc.gsfc.nasa.gov/docs/xanadu/xspec/python/html/index.html}} to analyse the spectrum in the 0.2-2.0\,keV energy band and quote errors for a 90\% confidence range for the derived spectral parameters. Since neither the spectrum nor the flux changed significantly from one observation to the other, we fit all 
six spectra (from the two observations and three instruments)
together and tie the spectral parameters. We first fit a simple absorbed black-body model to the data together with the Tuebingen-Boulder ISM absorption model 
\citep[\texttt{tbabs*bbody};][]{2000ApJ...542..914W}. The corresponding spectra and residuals (\texttt{delchi}) are shown in Fig.~\ref{fig:spec} for both observations. We can notice from the pn data, which has a higher sensitivity, that an absorption feature is left at the high-energy end of the spectra. We then add an absorption edge component (\texttt{tbabs*bbody*edge}). The values of the spectral parameters and the corresponding observed and unabsorbed fluxes in the 0.2--2.0\,keV, as well as the bolometric fluxes are given in Table~\ref{tab:spec} for both observations and both models: without (1) and with (2) an absorption edge. The observed (0.2--2.0\,keV) and bolometric luminosities (in erg s$^{-1}$) are provided in the last two columns and are obtained by multiplying the flux values by a factor of 4.22 $\times 10^{50}$, which assumes a distance to NGC~300 of 1.88\,Mpc.
The extra absorption edge helped to remove the residuals in the spectrum and to decrease the chi-square by $\Delta\chi^2=52.6$ for one degree of freedom less (see Table~\ref{tab:spec}). The absorption depth parameter was frozen to 1 since it could not be properly constrained. The absorption edge could be consistent with an NVII edge at 0.66 keV as it is also observed for M101 X-1 \citep{Kong2004}. The addition of this absorption edge also modified the spectral parameter values and reduced the bolometric luminosity to 2.7$\times 10^{38}$ erg s$^{-1}$, which is then compatible with the Eddington limit for 1.4\,M$_\odot$ white dwarf compact object.

Besides a black-body model, we also fitted the data with a disk black-body (\texttt{tbabs*diskbb}) used to characterize accretion disks consisting of multi-color black-body components, and with a thermal bremsstrahlung model (\texttt{tbabs*bremss}). However, for the bremsstrahlung model, it was not possible to constrain the energy of an absorption edge. The disk black-body models and even more extremely the bremsstrahlung model provide much higher bolometric fluxes/luminosities that can only be explained by super-Eddington accretion onto a  neutron star or a more massive black hole. The bolometric fluxes are however affected by large uncertainties since they are derived from supersoft spectra measured in a narrow energy range (0.2-2.0\,keV). For the black-body model, using a simple photoelectric absorption \texttt{phabs} model or the Tuebingen-Boulder ISM absorption model (\texttt{tbabs}) doesn't change spectral parameters significantly. On the other hand, for the disk black-body model with absorption edge, the energy of the edge could only be constrained with the \texttt{tbabs} model.

\begin{table*}	
\centering
	\caption{Spectral parameters and errors derived from the \xmm observations using a \texttt{tbabs*bbody} (1), a \texttt{tbabs*bbody*edge} (2), a \texttt{tbabs*diskbb} (3), a \texttt{tbabs*diskbb*edge} (4), a \texttt{tbabs*bremss} (5) model fit, with observed and unabsorbed fluxes in the 0.2--2.0\,keV band and bolometric fluxes. The corresponding observed and bolometric luminosities are shown in the last two columns.}
	\begin{tabular}{ccccccccccc} 
		\hline
		Model &  N$_\textrm{H}$ & kT& edge E & $\chi^2$/dof & $\chi^2_\textrm{red}$  & Flux & Unabs. Flux & Flux (bol.) & Lum. &  Lum (bol.) \\
		&  $\times 10^{20}$ &  &  & & & $\times 10^{-13}$ &  $\times 10^{-13}$ &  $\times 10^{-13}$&  $\times 10^{37}$ &  $\times 10^{38}$   \\
		&  ($\textrm{cm}^{-2}$) & (eV) & (keV) & & & (erg cm$^{-2}$ s$^{-1}$) & (erg cm$^{-2}$ s$^{-1}$) & (erg cm$^{-2}$ s$^{-1}$) & (erg  s$^{-1}$) & (erg s$^{-1}$) \\
		& & & & & & 0.2-2.0\,keV& 0.2-2.0\,keV & & 0.2-2.0\,keV & \\
		\hline
                  (1) &9.57$_{-0.92}^{+0.98}$&61$_{-1}^{+1}$& &386.4/292&1.32&1.0$_{-0.03}^{+0.006}$&7.8&14.1&4.3$_{-0.1}^{+0.02}$&5.9\\
                  (2) &6.68$_{-0.94}^{+1.04}$&72$_{-3}^{+3}$&0.64$_{-0.01}^{+0.02}$&333.8/291&1.15&1.1$_{-0.02}^{+0.008}$&4.2&6.5&4.5$_{-0.1}^{+0.03}$&2.7\\
                  (3) &10.91$_{-0.93}^{+0.98}$&69$_{-2}^{+2}$& &395.8/292&1.36&1.0$_{-0.06}^{+0.001}$&11.6&41.9&4.3$_{-0.2}^{+0.006}$&18\\
                  (4) &8.0$_{-0.92}^{+1.01}$&84$_{-4}^{+4}$&0.64$_{-0.01}^{+0.02}$&337.5/291&1.16&1.1$_{-0.09}^{+0.0008}$&6.1&16.6&4.5$_{-0.4}^{+0.003}$&7.0\\
                  (5) &12.51$_{-0.95}^{+1.0}$&92$_{-3}^{+3}$& &409.5/292&1.4&1.0$_{-0.04}^{+0.003}$&19.2&241.0&4.3$_{-0.2}^{+0.01}$&100\\
                 \hline		
	\end{tabular} 
	\label{tab:spec} 
\end{table*}

\section{Comparison with previous observations}
\label{sec:compar}

The source appeared as a bright supersoft X-ray source in two previous \xmm observations performed on the 26 December 2000 (obsID: 0112800201) and 1st of January 2001 (obsID: 0112800101)  for 37\,ks and 47\,ks respectively.
We fitted the spectrum again using more recent calibration files, with a simple absorbed black-body emission. We found temperatures of kT = 67$_{-7}^{+8}$ eV and kT = 77$_{-8}^{+9}$ eV, and hydrogen column densities N$_\textrm{H}$=10.06$_{-3.41}^{+3.97}$ and 4.01$_{-2.38}^{+2.86} \times 10^{20}$ $\textrm{cm}^{-2}$ for both observations, which are marginally compatible, considering uncertainties, from the ones derived in \cite{Kong2003}. The corresponding luminosities in the 0.2-2.0\,keV band are L$_\textrm{X}$=3.5$_{-0.9}^{+0.005}$ and 2.3$_{-0.3}^{+0.08} \times 10^{37}$ \ergs, and the bolometric luminosities are L$_\textrm{bol}$=3.3$_{-1.5}^{+4.5}$ and 0.75$_{-0.27}^{+0.56} \times 10^{38}$ \ergs, which are a factor of a few lower than what was reported previously \citep[15 and $1.9 \times 10^{38}$ erg s$^{-1}$,][]{Kong2003}.

The source was observed in outburst again in 2008 by Swift on the 20th of May and 4th of June 2008 \citep{Kong2008} which triggered a Chandra observation on the 8th of July 2008. The Chandra net count rate in the 0.2-2.0\,keV band was of $7.82 \times 10^{-3}$ cts s$^{-1}$, leading to L$_\textrm{0.2--2.0 keV}$ = $2.7 \times 10^{37}$ erg  s$^{-1}$ and L$_\textrm{bol}$ = $1.1 \times 10^{38}$ erg s$^{-1}$\, using the \texttt{CIAO 4.9 srcflux} tool\footnotemark[4]. 
Finally \cite{Kong2003} reported the source in outburst as well on the 26th of May 1992 in a ROSAT observation. Taking the count rate of $6.3\times 10^{-3}$ cts s$^{-1}$ from the ROSAT catalogue, we derived absorbed and unabsorbed luminosities  of $5\times 10^{37}$ erg  s$^{-1}$ and $3.9\times 10^{38}$ erg  s$^{-1}$ respectively\footnote{assuming a black-body spectrum with kT=65\,eV and N$_\textrm{H}$=10 $\times 10^{20}$ $\textrm{cm}^{-2}$} using the \texttt{WebPIMMS}\footnote{\url{https://heasarc.gsfc.nasa.gov/cgi-bin/Tools/w3pimms/w3pimms.pl}} tool. However, the source was not detected in four other pointed observations performed between 1991 and 1997 (one  with the PSPC and three with the HRI detector).

Table~\ref{tab:ul} provides the upper limit fluxes and luminosities\footnotemark[4], in the 0.2-2.0\,keV band, for the ROSAT, Chandra and \xmm  (EPIC pn instrument) observations where the source was not detected. The Chandra upper limit fluxes are derived from the  \texttt{CIAO 4.9 srcflux} tool, while for \xmm we extracted the upper limit count rates from the pn sensitivity maps and converted to fluxes\footnotemark[4].  ROSAT upper limits are given for a 3$\sigma$ confidence.

\begin{figure}
   \centering
   \resizebox{\hsize}{!}{\includegraphics{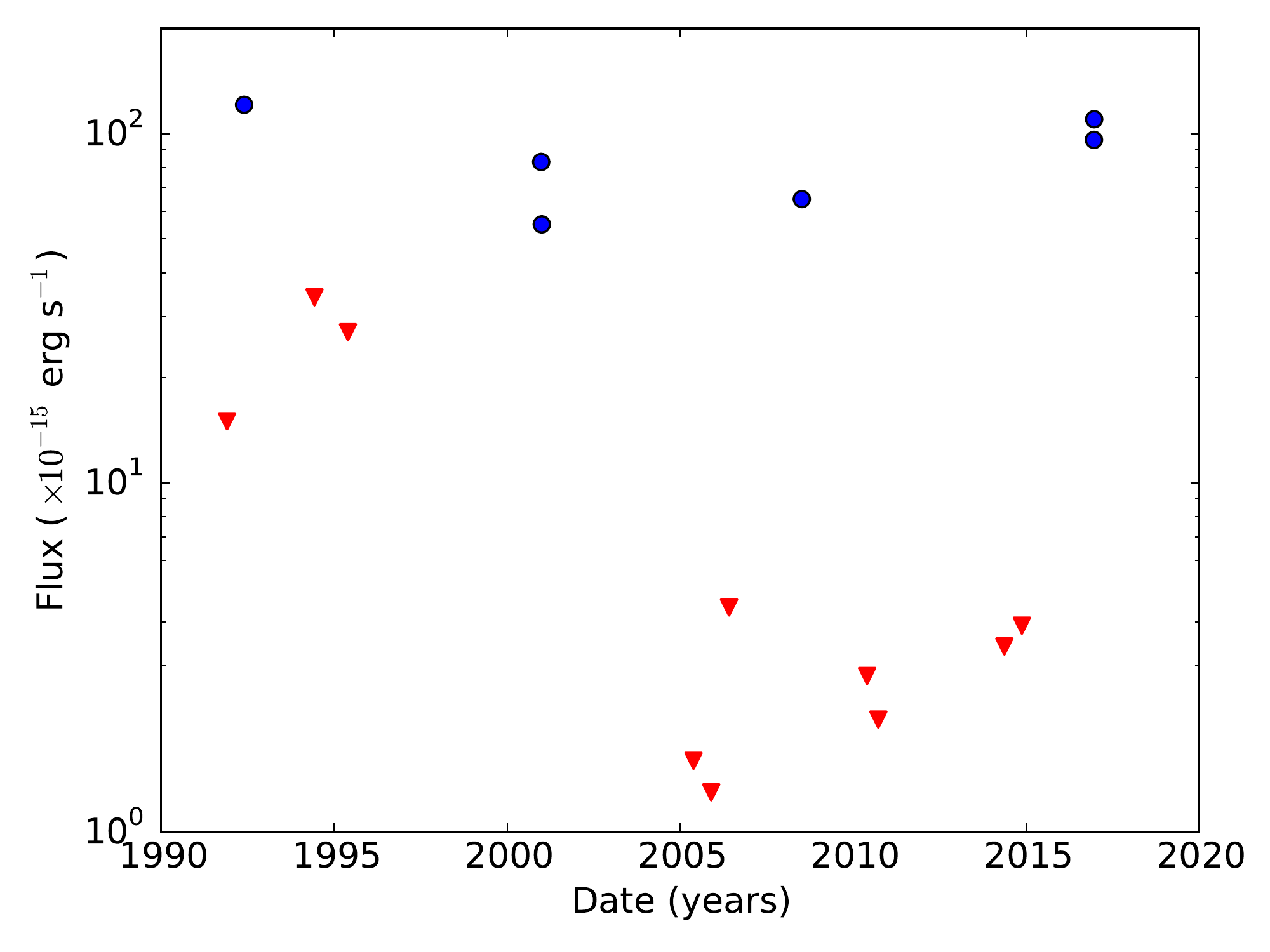}}
   \caption{Absorbed measured fluxes (blue dots, from Table~\ref{tab:spec} and Sec.~\ref{sec:compar}) and upper limits (red triangles, from Table~\ref{tab:ul}), in the 0.2-2.0\,keV band, from the ROSAT, \xmm and Chandra observations.}
              \label{fig:long_fluxes}             
\end{figure}

Figure~\ref{fig:long_fluxes} summarizes the absorbed measured fluxes (blue dots, from Table~\ref{tab:spec} and Sec.~\ref{sec:compar}) and upper limits (red triangles, from Table~\ref{tab:ul}) in the 0.2-2.0\,keV band, from ROSAT, \xmm and Chandra observations. From the highest measured fluxes from 1992 or 2016 to the lowest values for the upper limits from 2005, the flux varied by at least a factor of $\sim$80-90, making this source extremely variable.

The long-term light curve of NGC~300 SSS$_1$ may suggest that the source has outburst cycles of about 8 years (also $\sim$4 years is not excluded) since it was observed with a high luminosity in 1992, 2000, 2008 and 2016. More regular observations of the galaxy, with a highly sensitive instrument, would be necessary to confirm/infirm this prediction.

\begin{table*}
	\centering
\caption{Upper limit for fluxes and luminosities (absorbed and unabsorbed), in the ROSAT, Chandra and \xmm observations respectively where the source was not detected. The ROSAT HRI observation from 1997 is not reported due to the low exposure ($\sim$5.6\,ks).}

\begin{tabular}{cccccc} 
\hline
obsid & date & Flux (0.2-2.0 keV) & Lum. (0.2-2.0 keV) & Unabs. Flux (0.2-2.0 keV) & Unabs.  Lum (0.2-2.0 keV)\\
 & & $\times10^{-15}$ & $\times10^{36}$& $\times10^{-14}$ & $\times10^{37}$\\
 & & (erg cm$^{-2}$ s$^{-1}$) & (erg s$^{-1}$)& (erg cm$^{-2}$ s$^{-1}$) & (erg s$^{-1}$) \\
\hline
600025p-0  & 28-11-1991 &< 15.0 & < 6.3 & < 11.3& < 4.8 \\
600621h  &  08-06-1994 &< 34.5 & < 14.5& < 26.1 & < 11.0\\
600621h-1  & 27-05-1995 &< 27.5 & < 11.6 & < 20.8 & < 8.8 \\
\hline
7072 & 01-06-2006 & < 4.4& < 1.8& < 3.1 & < 1.3\\
12238 & 24-09-2010 & < 2.1 & < 0.9 & < 0.8 & < 0.3\\
16028 & 16-05-2014 & < 3.4 & < 1.4 & < 1.3 & < 0.6\\
16029 & 17-11-2014 & < 3.9 & < 1.7 & < 1.5 & < 0.7\\
\hline
0305860401 & 22-05-2005 & < 1.6 & < 0.7 & < 1.2 & < 0.5\\
0305860301 & 25-11-2005 & < 1.3 & < 0.6 & <  1.0 & < 0.4\\
0656780401 & 28-05-2010 & < 2.8 & < 1.2 & <  2.1 & < 0.9\\
\hline

\label{tab:ul}

\end{tabular}
\end{table*}

\section{Discussion}
\label{sec:discuss}
Nearby galaxies are hosting luminous supersoft X-ray sources with luminosities that can either be explained by steady nuclear burning of hydrogen accreted onto white dwarf surfaces (in the range of $10^{36}-2\times10^{38}$ \ergs) or accretion onto neutron stars or black holes (for luminosities $>2\times10^{38}$ \ergs). The bolometric luminosities derived for SSS$_1$ are extrapolated from a narrow energy range of 0.2$-$2\,keV and are subject to uncertainties, hence the nature of the compact object also cannot be ascertained clearly.

If the compact object is a white dwarf, one could be dealing with a nova, which undergoes a supersoft X-ray phase (delayed by several days/weeks with respect to the optical outburst). Multiple quasi-periodic outbursts would then suggest a recurrent nova.
We review here some examples of recurrent novae for which bright supersoft X-ray emission has been observed. RS\,Oph is a Galactic symbiotic recurrent nova that was observed in the X-rays as a supersoft source from week 6 to 10 after the optical outburst, with a luminosity close to the Eddington limit for a mass of at least 1.4\,M$_{\odot}$ \citep{Nelson2008}. The outburst events are described to be caused by a thermonuclear runaway at the white dwarf surface that are triggered by accretion disc instabilities \citep{Bollimpalli2018}. For another Galactic recurrent nova V745 Sco, the supersoft phase was detected only 4 days after the optical outburst of February 2014 and lasted for only 2 days, being the fastest evolving super-soft source phase yet discovered \citep{Page2015}. Both emission lines and absorption edges were found on the top of a black-body spectrum. Several recurrent novae have been found in the Andromeda galaxy as well, one of which being RX\,J0045.4+4154, which underwent in 2013 a SSS phase just 5 days after the optical outburst and lasted for only 12 days \citep{Tang2014}. The recurrence time between the outbursts is only 1 year for that source, while the bolometric luminosity, which depends on the fitted spectral model, is > $1\times10^{38}$ \ergs. Based on the observed recurrence time, the duration and effective temperature of the supersoft phase, the WD mass could be constrained to be >1.3\,M$_{\odot}$. These examples show that recurrent novae are able to explain the transient luminous supersoft X-ray emission observed in NGC~300 SSS$_{1}$. The relatively high temperature of $\sim70$\,eV derived from the X-ray spectra would then suggest a WD with mass $\gtrapprox$1.2\,M$_{\odot}$ \citep{2013ApJ...777..136W}.

Supersoft X-ray emission may also arise from systems containing black holes or neutron stars. The former is the preferred hypothesis for bolometric luminosities that are well above the Eddington limit for a 1.4\,M$_{\odot}$ compact object, although ultraluminous X-ray sources containing neutron stars have been confirmed in the last years thanks to the detections of pulsations \citep{Bachetti2014,Fuerst2016,Israel2017b,Israel2017a,Carpano2018}.
It has also been suggested that the supersoft X-ray emission may arise from massive outflows above the disk forming an optically thick photosphere which could shield the inner disk emission when viewed at high inclination angle (see \cite{Jin2011}, and references therein, for M81 ULS and \cite{Soria2016} for M101 ULS). This would mean that supersoft and normal ULXs are the same objects viewed at different viewing angles. 
Some sources in M~101 and NGC~247, on the other hand, have been observed in some epochs in their ULS state and in others in their ULX state, which can be interpreted as a change in the accretion rate or in the viewing angle, due to disk precession. The transition between ULSs (only soft component) and ULXs (hard tail) could occur at black-body temperatures of $\sim$150\,eV \citep{Urquhart2016}.
A powerful wind characterized by blue-shifted emission and absorption lines was found by \cite{Pinto2017} in the \xmm RGS spectrum of the ULX/ULS source in NGC~55. Strong variability on time scale of $\sim$1\,ks was observed for M~101 ULS,  with an rms variability of $\sim30-100\%$, which is higher than what is observed for stellar-mass black holes in their canonical state, but quite common for ULXs \citep[and references therein]{Soria2016}. This has been interpreted as due to a variable obscuration of the harder central regions by a clumpy disk wind. No periodic modulation was observed neither for the SSS in M101 \citep{Soria2016} nor for the one in M81 \citep{Liu2008}. A potential recurrence time for high states of around 160 and 190\,d or possibly around 45\,d, with durations exceeding 18\,d was reported by \cite{Mukai2005}  for  M101~ULX-1, that could be caused by a radiation-induced warping of the accretion disk.

Instead of an optically thick wind, \cite{Gu2016} proposed that the soft black-body component of ULSs could come from a geometrically thick accretion disk. In this model both the accretion rate should be high ($\dot{M}>30\,\dot{M}_\textrm{Edd}$, where $\dot{M}_\textrm{Edd}$ is the Eddington accretion rate) and the inclination angle not too small (>25$^\circ$), and only the soft photons from the outer thin disk and the outer photosphere of the thick disk can be viewed. The radius at which the disk switches from geometrically thick to thin is called the transition radius. This model would naturally explain the large black-body radius extracted from the L$_\textrm{bol}$ = 4$\pi R_\textrm{bb}^2 \sigma T_\textrm{bb}^{4}$ formula.
In the case of SSS$_{1}$, assuming the spectral parameters of the best fitted model (2) in Table~\ref{tab:spec}, the photosphere would have a radius (i.e. a transition radius) of $8.8 \times10^3$\,km. Note that if the compact object is a white dwarf instead of a neutron star or black hole, the black-body radius extracted from the above formula would be fully consistent with a white dwarf radius.
With respect to the optically thick outflow model, this model of a geometrically thick accretion disk requires lower accretion rates.

The presence of intermediate mass ($100-1000 \textrm{\,M}_{\odot}$) black holes accreting at sub-Eddington luminosities is also quoted in the literature to explain ULS's X-ray emission. In that case the accretion occurs in a thin disk which is optically thick, with the supersoft flux originating from its inner edge at a  radius of the last stable orbit. However, such massive black hole accretors are not expected to be in their high/soft state at those luminosities that are well below their Eddington limit \citep{Urquhart2016}. For NGC~300 SSS$_1$  the moderate observed luminosity doesn't require the presence of an intermediate mass black hole, and the previously mentioned models are preferred.

\section{Conclusions}
\label{sec:conclu}
We report here the detection of a new outburst of the luminous supersoft source SSS$_{1}$ in NGC~300, observed twice between the 17th and 20th of December 2016 with \xmm. The bolometric luminosity  is superior or equal to $3\times10^{38}$ \ergs\ (depending on the spectral model) and is marginally consistent with a 1.4\,M$_{\odot}$ white dwarf accreting at Eddington luminosities. A periodic modulation of $4.68\pm0.26$\,h is measured in the first observation, which is still compatible with the period measured in Jan. 2001 ($5.7\pm1.1$\,h), affected by large uncertainties. This period could be associated with an orbital modulation, although the signal strength is highly variable. The source was observed in outburst 4 times in the last 30 years (in 1992, 2000, 2008 and 2016) suggesting a possible recurrence period of about 8 (or 4) years.

The system could be associated with a recurrent nova, since those systems undergo supersoft X-ray phases with luminosities close to the Eddington limit for a massive white dwarf and because the outbursts may occur periodically. Another possible explanation would be a ULX binary system viewed at high inclination angle, where the soft photons would come either from a disk outflow or from a geometrically thick disk. The presence of an intermediate mass black hole is less likely.

\section*{Acknowledgements}
This research is based on data obtained with \xmm, an ESA science mission with instruments and contributions
directly funded by ESA Member States and NASA, and makes use of data obtained from the Chandra, NASA/ESA Hubble Space Telescope and ROSAT Data Archive. The \xmm\ project is supported by the Bundesministerium f{\"u}r Wirtschaft und Technologie/Deutsches Zentrum f{\"u}r Luft- und Raumfahrt (BMWI/DLR, FKZ\,50OG1601) and the Max Planck Society.




\bibliographystyle{mnras}
\bibliography{paper} 

\bsp	
\label{lastpage}
\end{document}